**Technical Conference Paper**

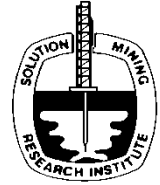

# Sounds good?

# Determination of a Gas/Brine Interface

# by an Acoustic Method

# at Manosque

**Benoit Brouard, Brouard Consulting, Paris, France**

**Pierre Bérest, Ecole Polytechnique, Palaiseau, France**
**Vincent de Greef, Ecole Polytechnique, Palaiseau, France**
**Gérard Gary, Ecole Polytechnique, Palaiseau, France**

**Jean-Paul Crabeil, Flodim, Manosque, France**

# SOUNDS GOOD? DETERMINATION OF A GAS/BRINE INTERFACE BY AN ACOUSTIC METHOD AT MANOSQUE

Benoît Brouard, Brouard Consulting, Paris, France

Pierre Bérest, Vincent de Greef, Gérard Gary, Ecole Polytechnique, France

Jean-Paul Crabeil, Flodim, Manosque, France

**Abstract**

Water hammers commonly are observed at wellheads and often are considered a potential hazard that should be avoided. Nevertheless, there are a few situations in which water hammers provide very valuable information about a well. A comprehensive data-acquisition and analysis system has been developed by Brouard Consulting and Ecole Polytechnique. One potential application of that system is determining the depth of an interface between two fluids. This application has been tested successfully at the Manosque field. It is demonstrated here how this low-cost and non-intrusive system can be accurate and allows practical, real-time measurements.

**Key words:** Well Logging, Instrumentation and Monitoring, Computer Modeling/Software, Leaching, MIT.

## 1. Introduction

In the solution-mining industry, there are several situations in which it is useful — or even necessary — to know precisely the final length of a hanging string, the depth of an interface between two fluids, or the location and size of a plug. Examples of such situations are given in Figure 1.

A damaged (or even broken) string often is not detected in real time. An interface-level measurement usually is taken only a couple of times during a Mechanical Integrity Test (MIT, tightness test using nitrogen), and, typically, only once a month when leaching. In all cases, a relatively costly logging operation is required to obtain the needed data. Subsequently, it is appealing to build a non-invasive system that would permit accurate determination of the final lengths of strings or of interface depths, especially **in real time** and **at low cost**.

## 2. Usefulness of water hammers

When the momentum of the fluid in a wellbore or other piping system is changed rapidly, a flow discontinuity is initiated locally, and a pressure wave is transmitted through the fluid. The pressure wave produced at shut–in travels the length of the well, is reflected at the bottom, and returns to the surface, where it is reflected again. The wave travels in the fluid at the speed of sound. This process continues until the wave is completely damped by frictional processes within the fluid and along the casing walls.

For many years, Brouard Consulting and Ecole Polytechnique have measured and studied water hammers and, more generally, oscillatory phenomena in storage wells and salt caverns (Bérest et al., 1996 and 1999). Because water hammers are relatively easy to measure and because the related pressure evolution depends on well properties, there are several circumstances in which it could be useful to analyze water hammers to back-calculate various well parameters.

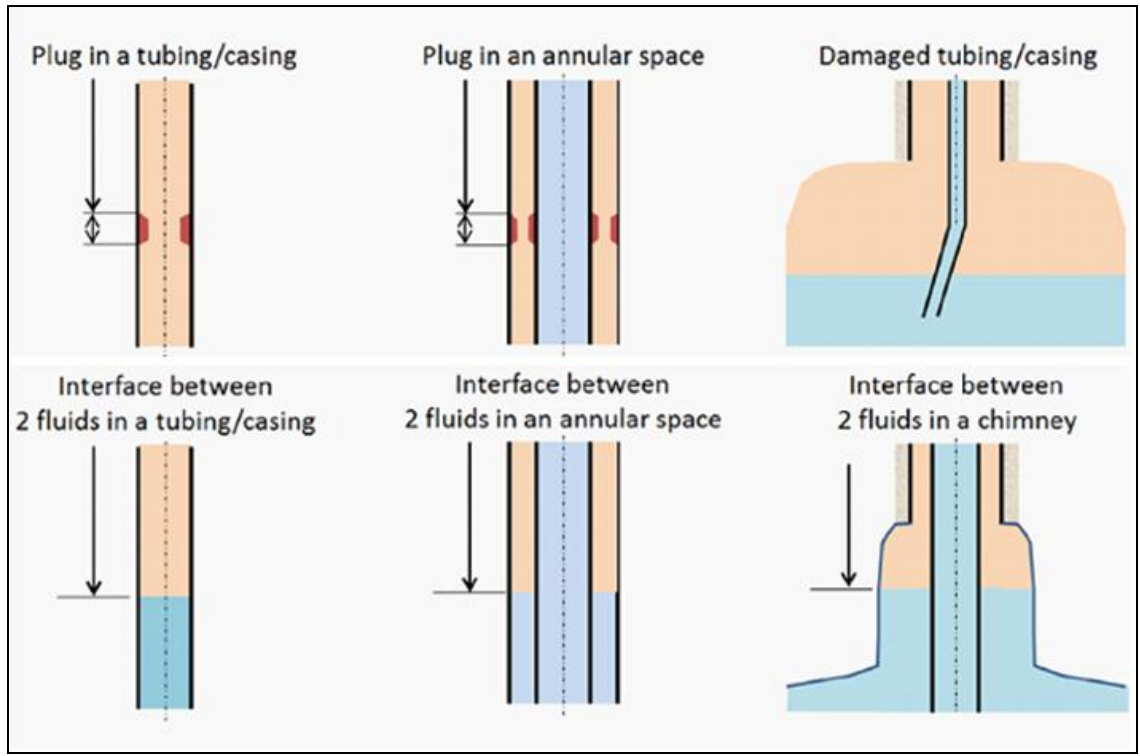

**Figure 1.** Examples of configurations in which water-hammer analysis could be helpful.

### 3. Background — Detection of damaged/lost strings at Vauvert

The detection of damaged strings using a water-hammer-based method was presented by Brouard et al. (2012). The main results of their study are given below.

#### *3.1 Vauvert case*

Arkema, now Kem One, produces brine at its Vauvert facility in southern France by solution mining at a depth ranging from 1900 m to 3000 m. The salt has been leached out between pairs or triples of wells since the 1970s. Connection between wells is obtained by hydrofracturing. Due to the complex geology and the great depths, salt production there is often difficult. Cavern creep closure is very fast, and many strings are lost during operation, leading to fast upward displacement of the cavern roof. Cavern-roof rise by hundreds of meters after a dozen years is typical. Because lost string cannot be detected from the wellhead, logging operations are necessary at least once a year for each operating well to determine the actual length of the well and the location of the cavern roof. Also, plugs periodically form in the casing, which can lead to very costly work-over operations.

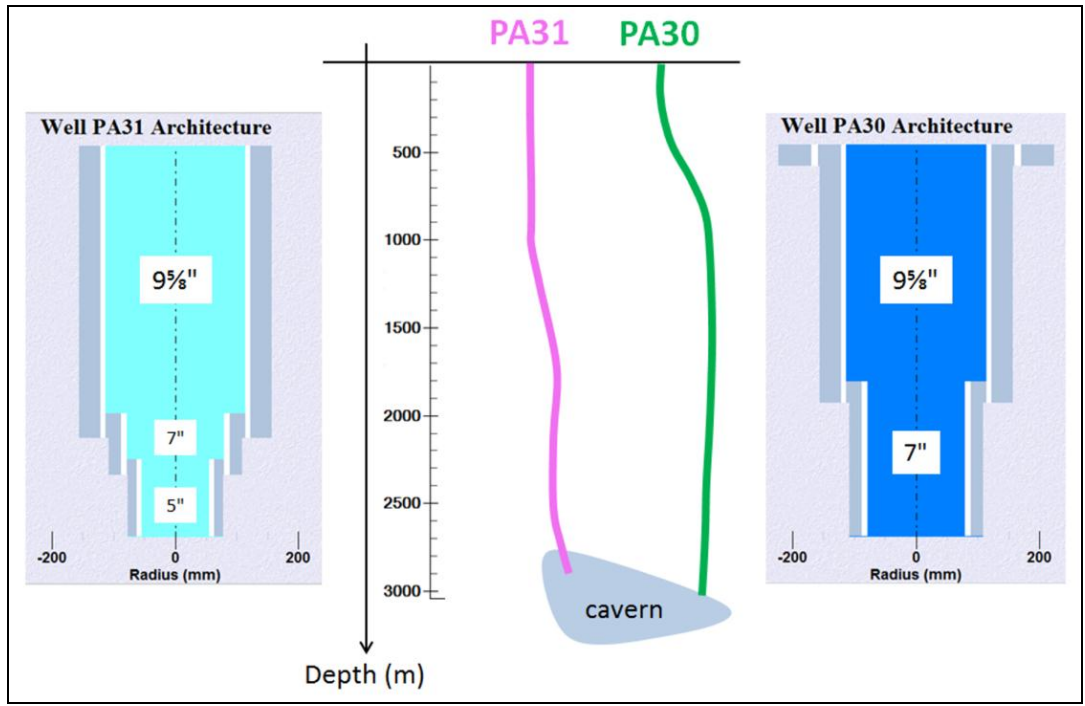

**Figure 2.** Schematic cross–section and completions of wells PA30–PA31 at the Vauvert facility.

### *3.2 Monitoring system*

In 2012, a permanent monitoring system was set up at the Vauvert facility by Brouard Consulting and Ecole Polytechnique. The installed system consists of careful data collection at two properly instrumented wells, the PA30-PA31 pair (Figure 2), followed by filtering, spectral analysis to enhance and identify meaningful pressure oscillations in the data and subsequent modeling to derive the final casing length as well as other possible information. A dedicated quartz pressure gauge was set up on well PA31 to allow accurate recording of water hammers. A set of high-pressure valves also was plugged at the wellhead. These serial valves are dedicated to the manual triggering of water hammers. The gauges are connected to the control room of the facility through buried cables (Figure 3). The dedicated data acquisition system in the control room records all data from the PA30-PA31 pairs — i.e., wellhead pressures (including from the quartz gauge), temperatures and flow rates.

A dedicated computer also was installed in the head office and permanently connected to the acquisition system in the control room through a power-line Ethernet network. Furthermore, this computer is connected to Internet, allowing remote control from anywhere outside the facility. This allows easy checking and updating of the system.

The system is able to record any significant water-hammer event that occurs. A pre-trigger allows wellhead pressures to be recorded a few seconds before the detection of any event, such as a water hammer triggered by a pump stop.

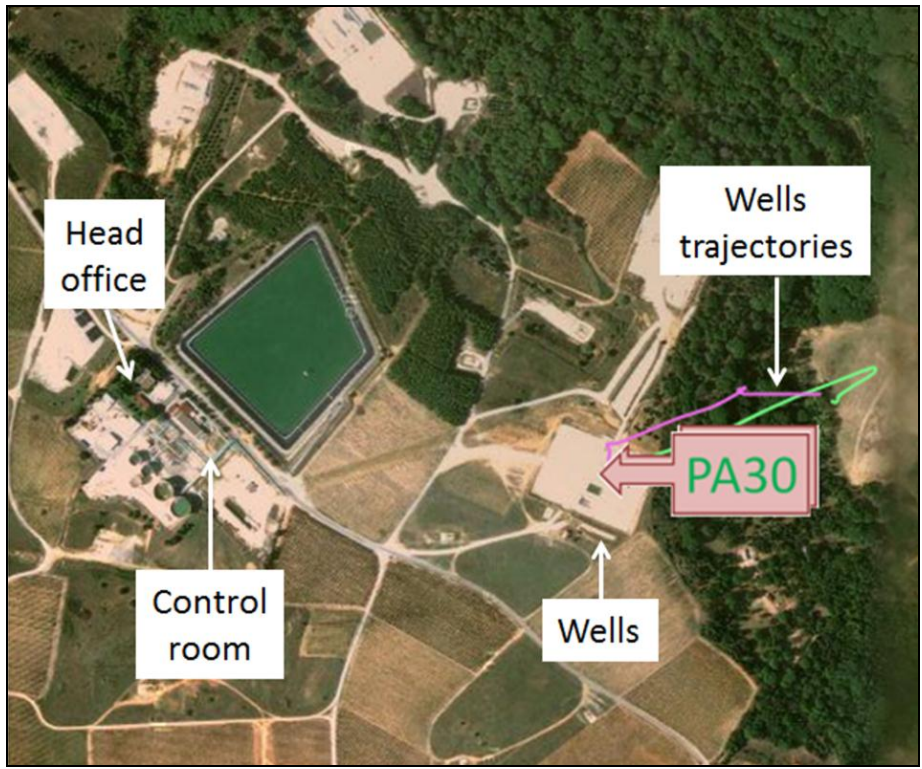

**Figure 3.** Overview of the Vauvert facility (Google Map).

*3.3 Data acquisition and analysis software*

The developed software allows the user to remotely control the acquisition system (Figure 3) to both display and analyze all acquired data. For each selected well, tables summarize all relevant recorded data and display pressure, temperature and flow rate evolutions.

All detected water hammers are controlled, and relevant signals are recorded in a database (Figure 4) and analyzed automatically. Significant events, such as a damaged or plugged casing, can be emailed automatically to a list of chosen recipients.

*3.4 Transient analysis*

A simulation example is provided by Figure 5, which provides a comparison between measured and computed pressure evolutions after a wellhead valve was left open for approximately 1-2 seconds. This simulation was performed by considering the initial length of the well before the start of leaching, which was significantly larger than the actual length at the time of the measurement. When this initial well completion is considered, the computed pressure evolution is clearly late compared to the measured pressure evolution. Furthermore, damping parameters are not optimized.

Figure 6 shows a comparison between measured and computed pressure evolutions after fitting. The final length of the well, celerities and damping coefficients in each part of the well were back-calculated. In this example, using the software developed for this project, it was found that the final depth of the casing was 2351.2 m MD (2335.4 TVD). A logging operation performed by Schlumberger a couple of weeks later found the final length of the casing at 2352.8 m MD — i.e., very close to the figure estimated previously.

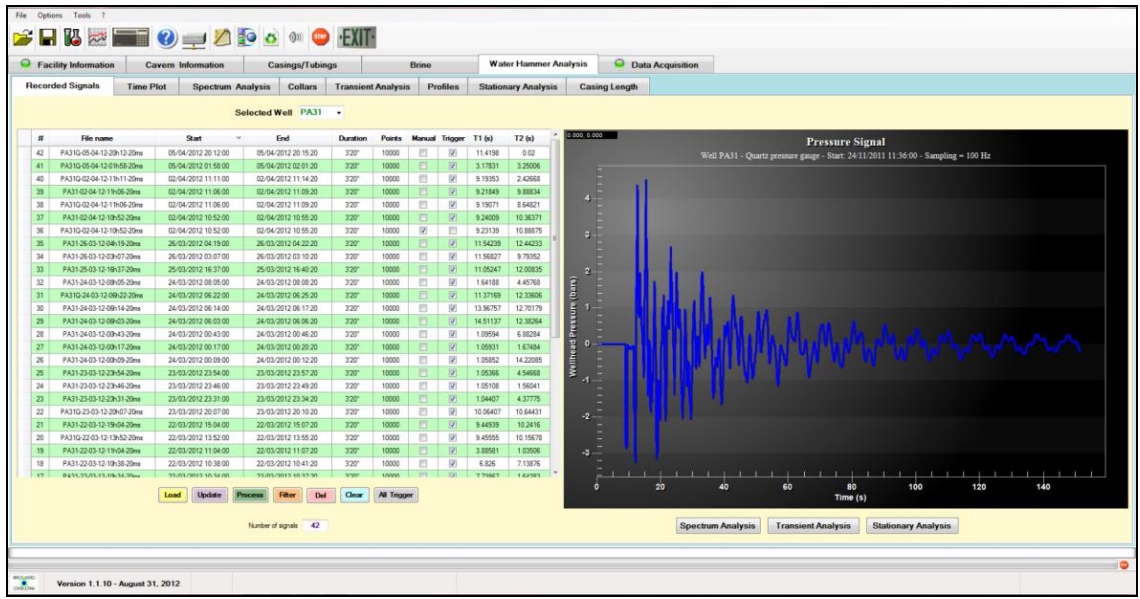

**Figure 4.** Screenshot of the PA31 water–hammer database.

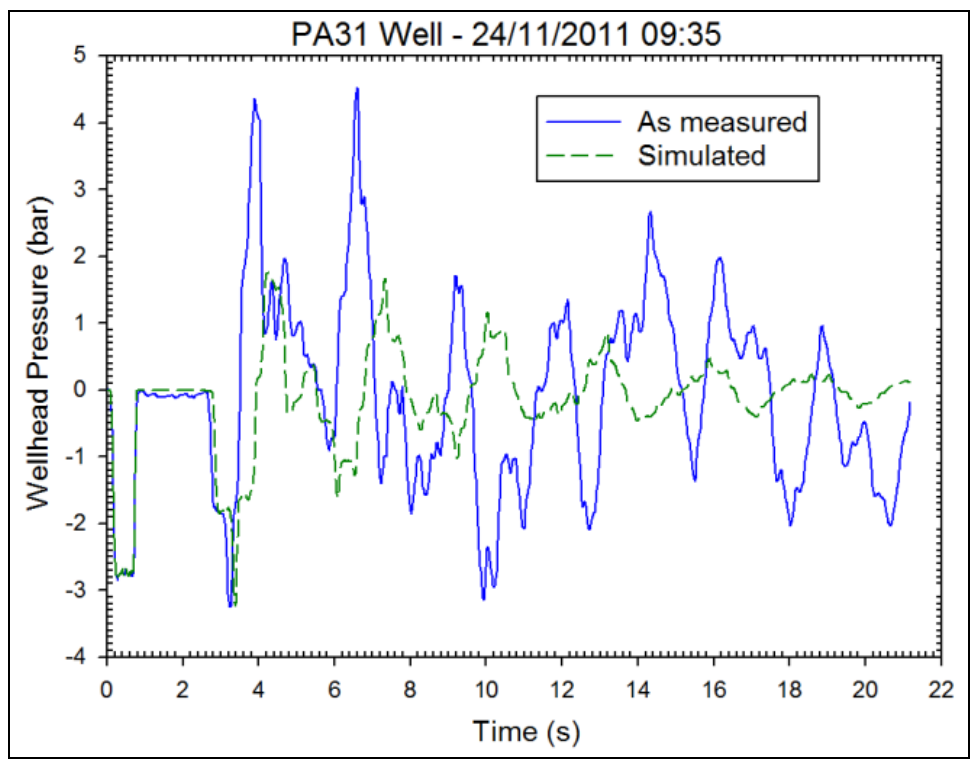

**Figure 5.** Comparison between measured and computed pressure evolutions at the Vauvert facility. Well properties were estimated from initial well completion.

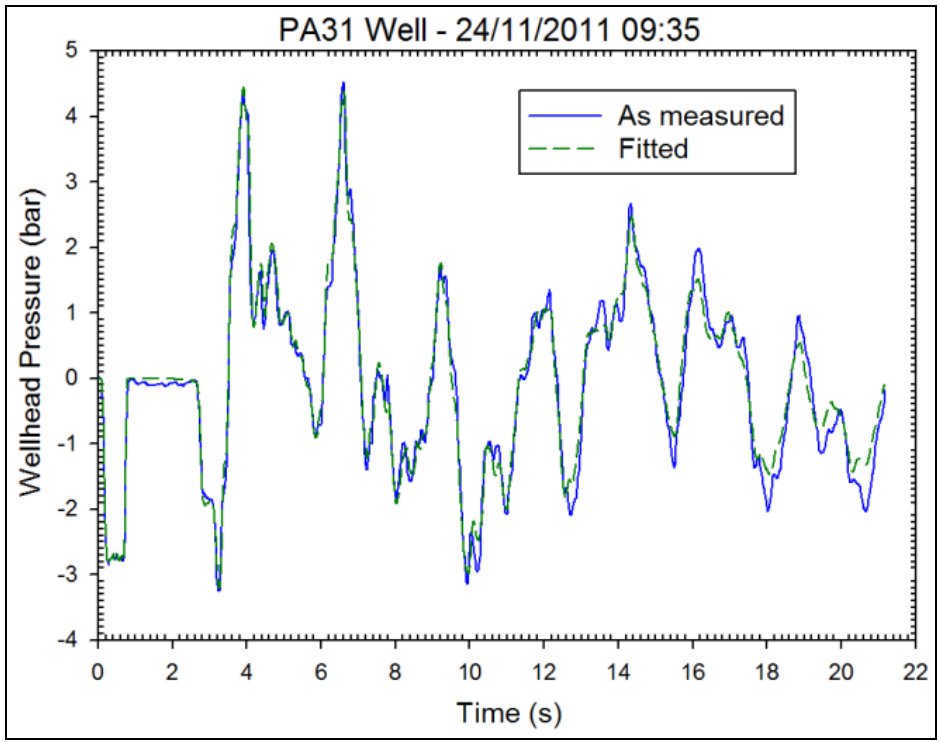

**Figure 6.** Comparison between measured and computed pressure evolutions after fitting at the Vauvert facility.

## 4. Nitrogen/Brine interface detection at Manosque

### *4.1 Well configuration*

When performing an MIT (Van Sambeek et al., 2005), it is usually necessary to measure the displacement of an interface between nitrogen and brine (Nitrogen Leak Test) or between two liquids (Liquid–Liquid Test) in a chimney. The same configuration can be found when leaching with a gas or liquid blanket (Figure 7).

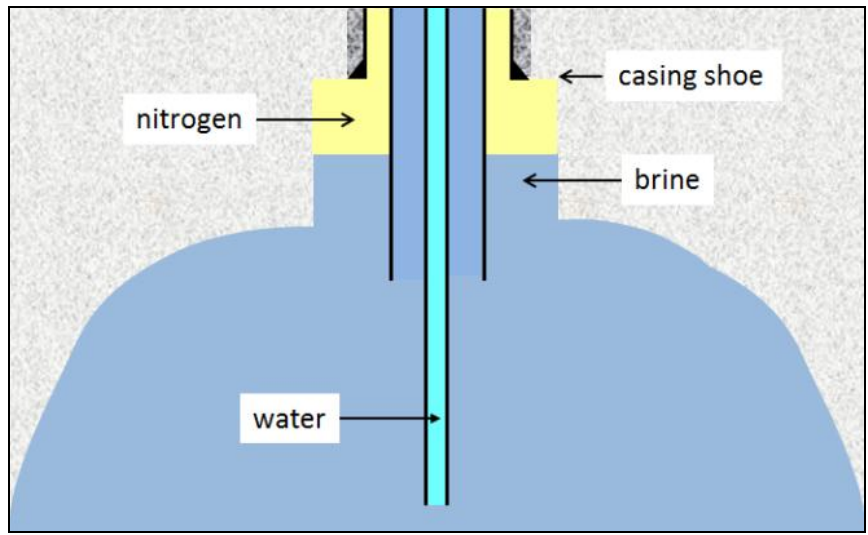

**Figure 7.** Typical configuration of a well during leaching.

**Manosque test**

Geostock operates a liquid hydrocarbon storage facility at Manosque (southeastern France), owned by Geosel-Manosque. This facility has 26 salt caverns that have been leached out since the early 1960s. Crude oil, diesel oil, naphtha and kerosene are stored in this facility (de Laguérie and Durup, 1994; de Laguérie and Cambon, 2010).

A new cavern, referred to as TA, has been leached out for a few months using two wells, TA1 and TA2. Soft water currently is being injected into the TA1 well while saturated brine is being withdrawn from the TA2 well. A test was designed to determine if it were possible to find the depth of the nitrogen/brine interface in the chimney during leaching using only water hammers. The configuration of the TA1 well when the test was performed is shown in Figure 8. The cemented part of the well is composed of a water-filled 10-¾" casing and a cemented 13-⅜" casing; the narrow annular space (22 l/m) is filled with nitrogen. The casing-shoe depth is 603 m below ground level. The nitrogen/brine interface is located in an open hole with an initial drilled diameter of 17-½". The depth of the cavern roof is estimated to be 740 m. The precise shape of the chimney is unknown.

**Testing system**

The testing equipment was composed of a data acquisition system, a quartz pressure gauge from Kistler and a gas gun from Echometer (Figure 9). This gun assembly consists of a gun and microphone together; it allows a quick and confined pressure pulse to be generated. An embedded microphone converts acoustic pressure pulses into an electrical signal. The gun and microphone normally are supplied together as a unit. This type of gas gun is used commonly in the petroleum industry to determine the downhole pressure when the annular space is filled with gas. The method consists of measuring the reflection on joint collars and on the gas/liquid interface to back-calculate gas column weight. One of the objectives of Manosque test was to test this type of equipment and to compare it to a quartz gauge in the particular context of a salt cavern.

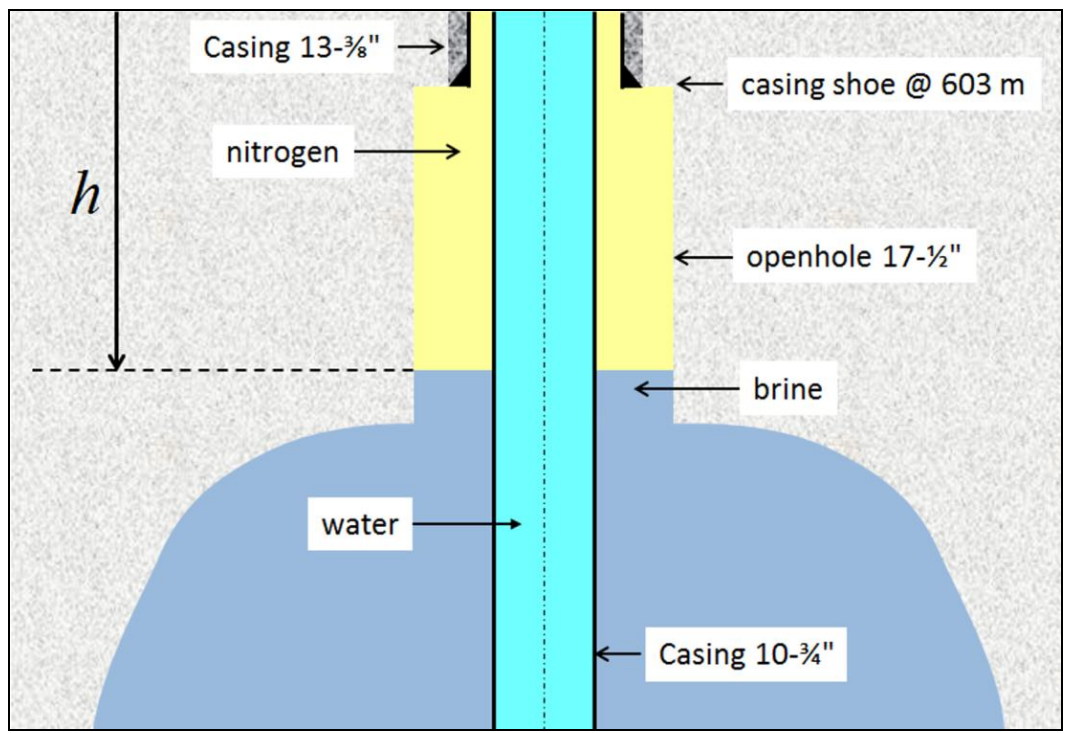

**Figure 8.** Basic configuration of the Manosque TA1 well.

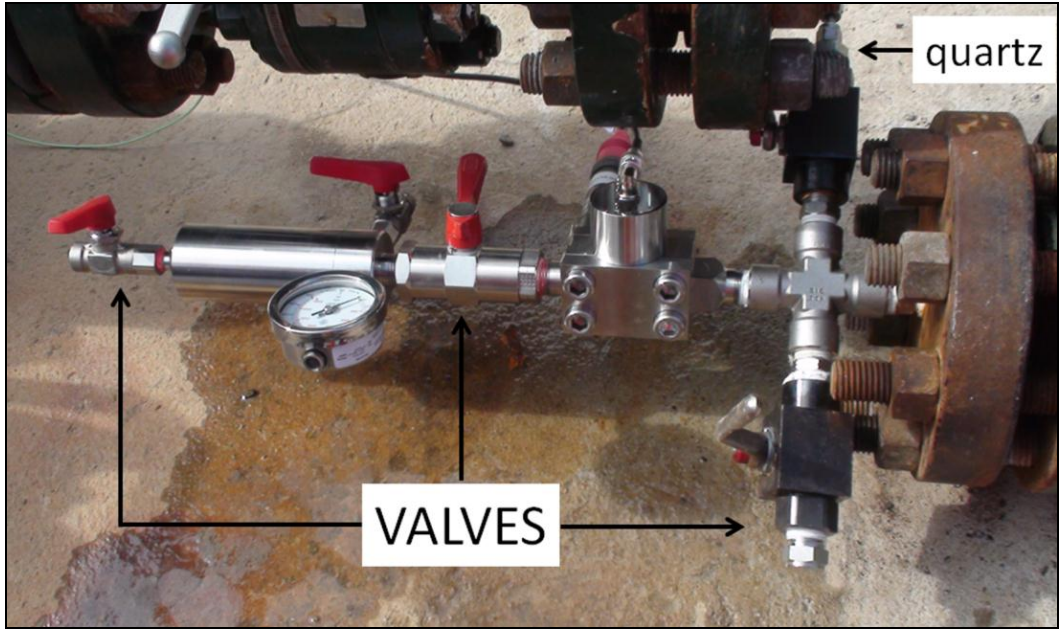

**Figure 9.** Testing equipment connected to the 10-¾"×13-⅜" annulus.

### *4.2 Quick signal analysis — Uniform celerity*

An example of a recorded raw pressure evolution at the wellhead during a manual water-hammer event (a wellhead valve was left open for approximately 1 second) is given in Figure 10 and Figure 11 using the quartz gauge and in Figure 12 using the microphone, respectively. Wave reflections at the casing shoe, at the interface, and then at the cavern roof can be observed very clearly on the raw signal from the quartz gauge. The signal from the microphone allows the first reflection at the casing shoe and on the interface to be determined, but the signal-to-noise ratio is smaller compared to the quartz gauge, and all low frequencies vanish, preventing any accurate fitting to get a more accurate value of the interface depth and also to get more information about the chimney (see Section 4.5).

During the first seconds, high frequency oscillations due to reflections on 10-¾" joint collars can be observed clearly using both gauges. The frequency of reflections on joint collars can be determined accurately from relevant power spectrums (Figure 13); this frequency is $F_{jc}$ = 37.79 Hz ± 0.01 Hz. As spatial distribution of the collars is known, an average tube length in the cemented part of the well equal to $L_{tub}$ = 9.56 m ± 0.02 m can be calculated; thus, the average celerity of sound in the entire nitrogen column can be estimated to be

$$\bar{c} = F_{jc} \times L_{tub} = 37.79 \times 9.56 = 361.3 \text{ m/s} \tag{1}$$

and the uncertainty is

$$\Delta \bar{c} = \bar{c} \times \left( \Delta F_{jc} / F_{cj} + \Delta L_{tub} / L_{tub} \right) \approx \pm\ 0.8 \text{ m/s} \tag{2}$$

High and low frequencies in both signals can be withdrawn using a band-pass filter, which allows easier detection of casing-shoe and interface reflections (Figure 14 and Figure 15). From Figure 15, it can be calculated that the travel time of the wave from the casing shoe, at a 603-m depth, to the nitrogen/brine interface is equal to $t_{ci}$ = 0.574 s ± 0.001 s. Therefore, the interface depth can be roughly estimated to be

$$h = H_{cs} + \left( t_{ci} \times \bar{c} \right)/2 = 603.0 + \left( 0.574 \times 361.3 \right)/2 = 706.7 \text{ m} \tag{3}$$

and the uncertainty is

$$\Delta h = h \times \left(\Delta t_{ci}/t_{ci} + \Delta \bar{c}/\bar{c}\right) \approx h \times \left(1{,}74 \times 10^{-3} + 2{.}21 \times 10^{-3}\right) \pm 2.8 \text{ m} \qquad (4)$$

Uncertainty on travel time, $\Delta t_{ci}$, comes mainly from sampling. The Echometer microphone sampling frequency is 4 kHz. The quartz gauge allows higher frequencies; a 10 kHz sampling was successfully tested, therefore it is expected that uncertainty $\Delta t_{ci}$ could be reduced.

The uncertainty relative to wave celerity, $\Delta \bar{c}$, is of the same order of magnitude. The gas gun from Echometer is provided with a software called TWM that calculates an average celerity of the travelling wave in the middle part of the well. It gave a 355.8 m/s average celerity and an estimation of interface depth at 696 m; i.e. significantly different from our former estimation.

### *4.3 Celerity correction*

Because the TA1-TA2 cavern is under active leaching, the brine temperature in the cavern is relatively cold compared to the temperature of the surrounding rock mass. A 18 °C temperature at the cavern roof was measured by Flodim in July 2010 (Figure 16). (The natural temperature of the salt may be estimated to be 37°C at this depth.) Well J is a nearby well of an idle oil-storage cavern (Gatelier et al., 2008). The average temperature of the entire nitrogen column during the water-hammer tests was probably of the order of 24°C, while the average nitrogen temperature in the chimney was close to 20°C. A 4 °C difference can lead to a reduction of wave celerity by 2.4 m/s in pure nitrogen [SMRI Toolbox]; thus, the interface depth should be corrected to be:

$$h = H_{cs} + \left(t_{ci} \times \bar{c}\right)/2 = 603.0 + (0.574 \times 358.9)/2 = 706.0 \text{ m} \qquad (5)$$

It also should be pointed out that the nitrogen column may contain some moisture, especially in the chimney part, which also can lower the sound-wave celerity slightly.

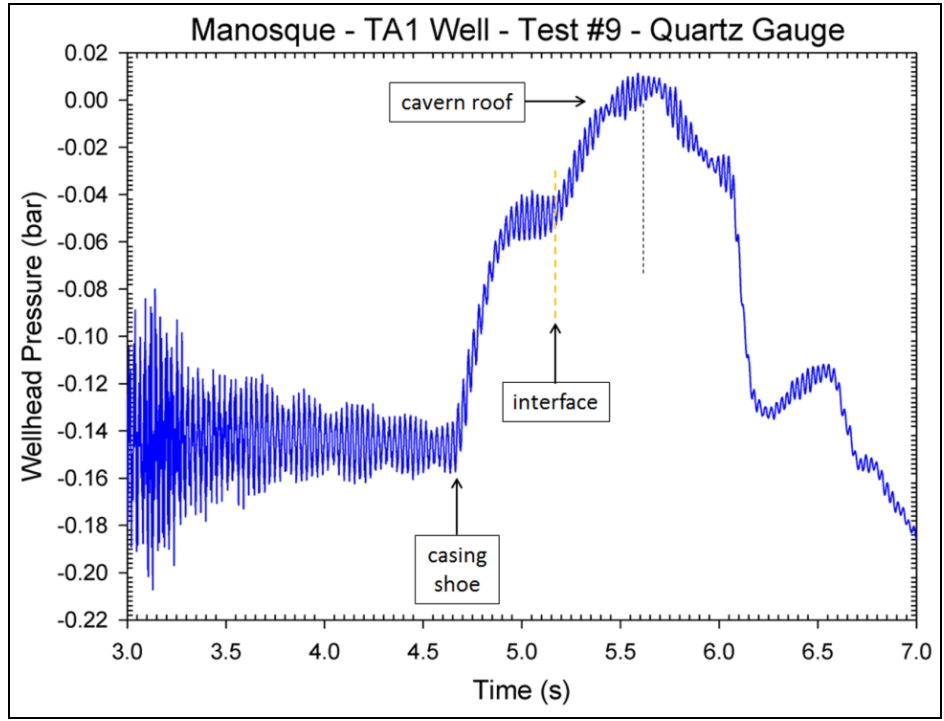

**Figure 10.** Close-up of measured pressure evolution.

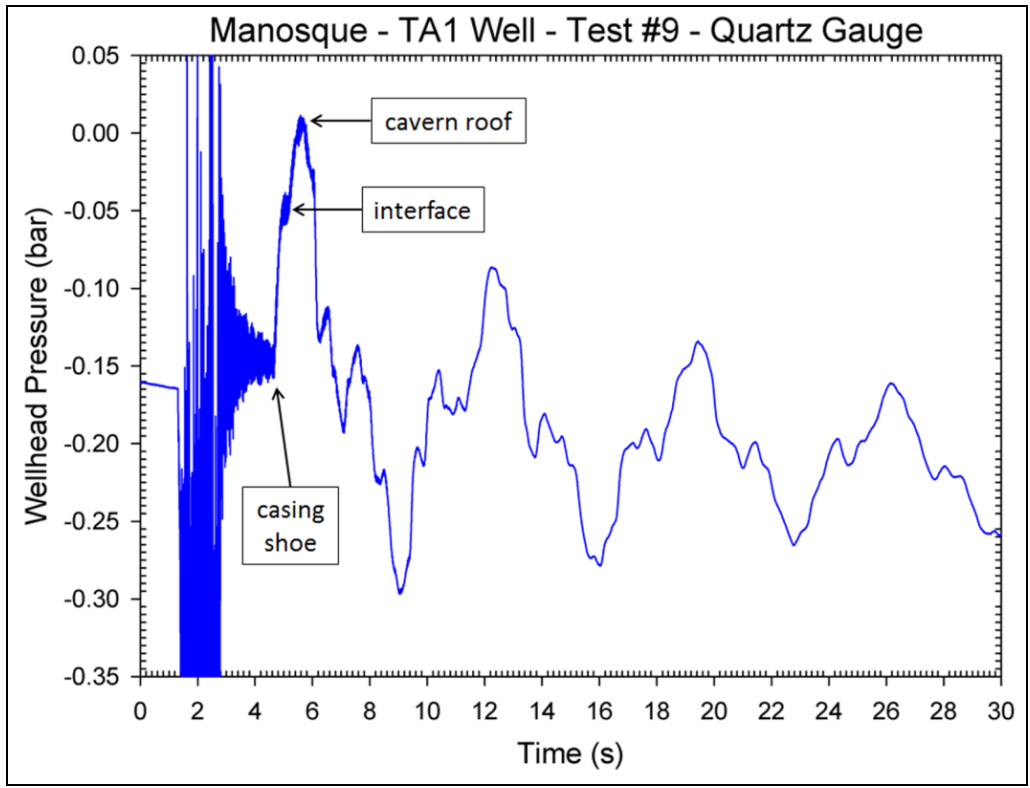

**Figure 11.** Measured pressure evolution using the quartz gauge.

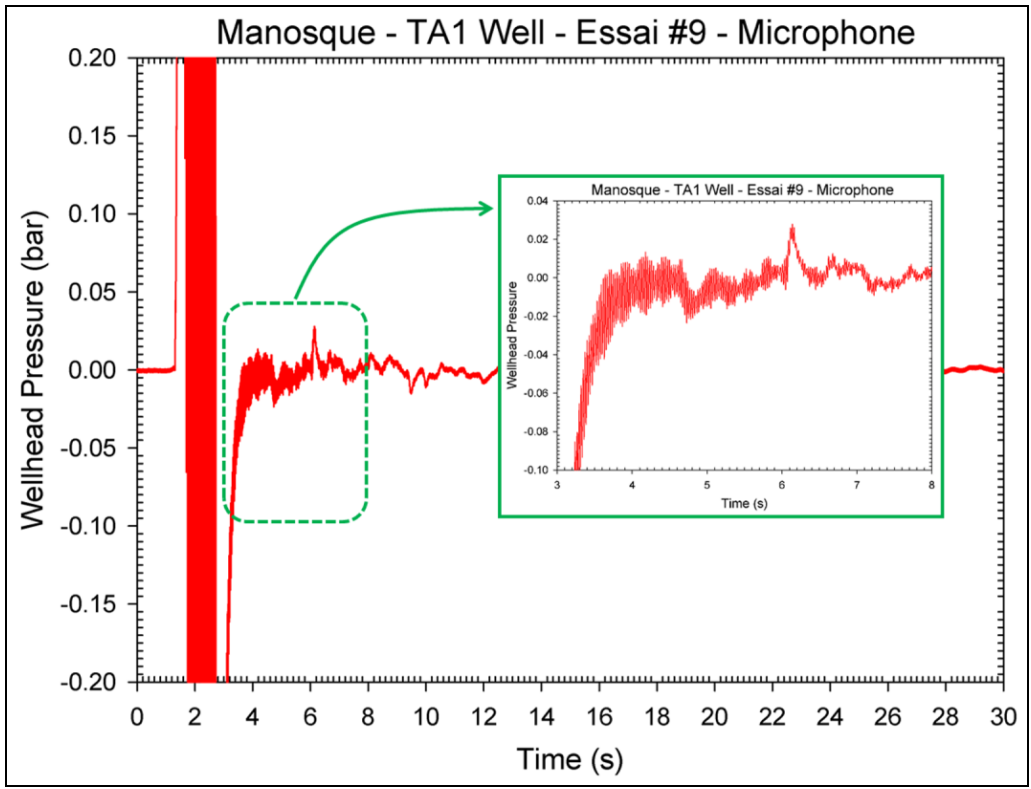

**Figure 12.** Measured pressure evolution using the microphone.

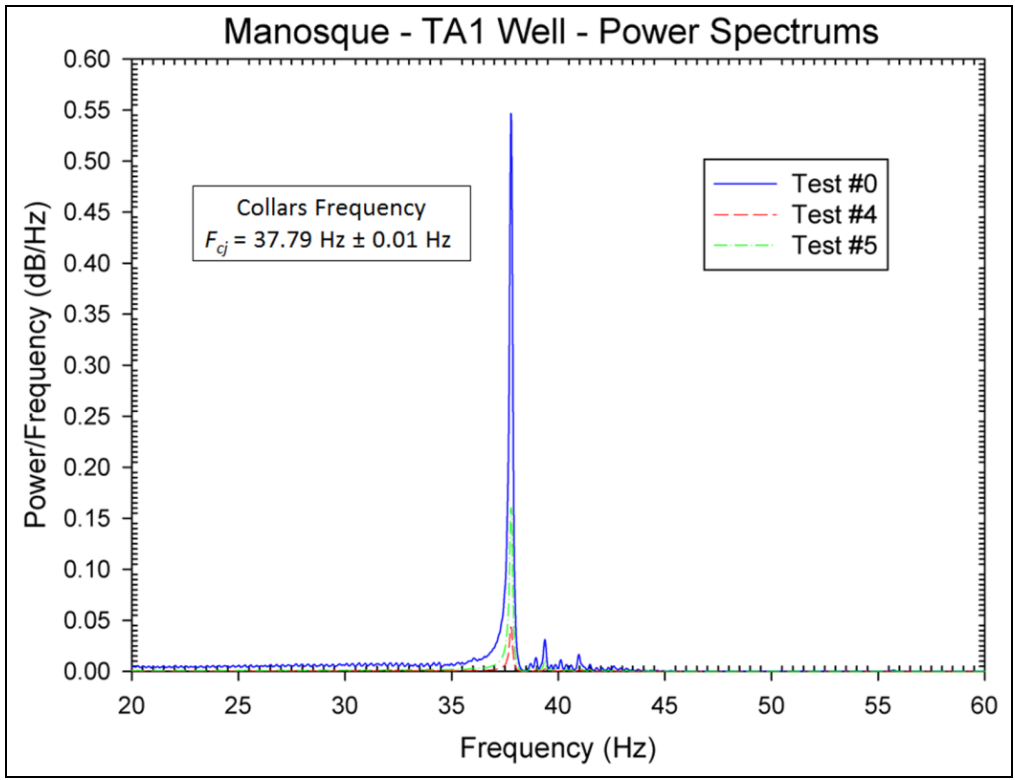

**Figure 13.** Comparison of power spectrums at Manosque TA1.

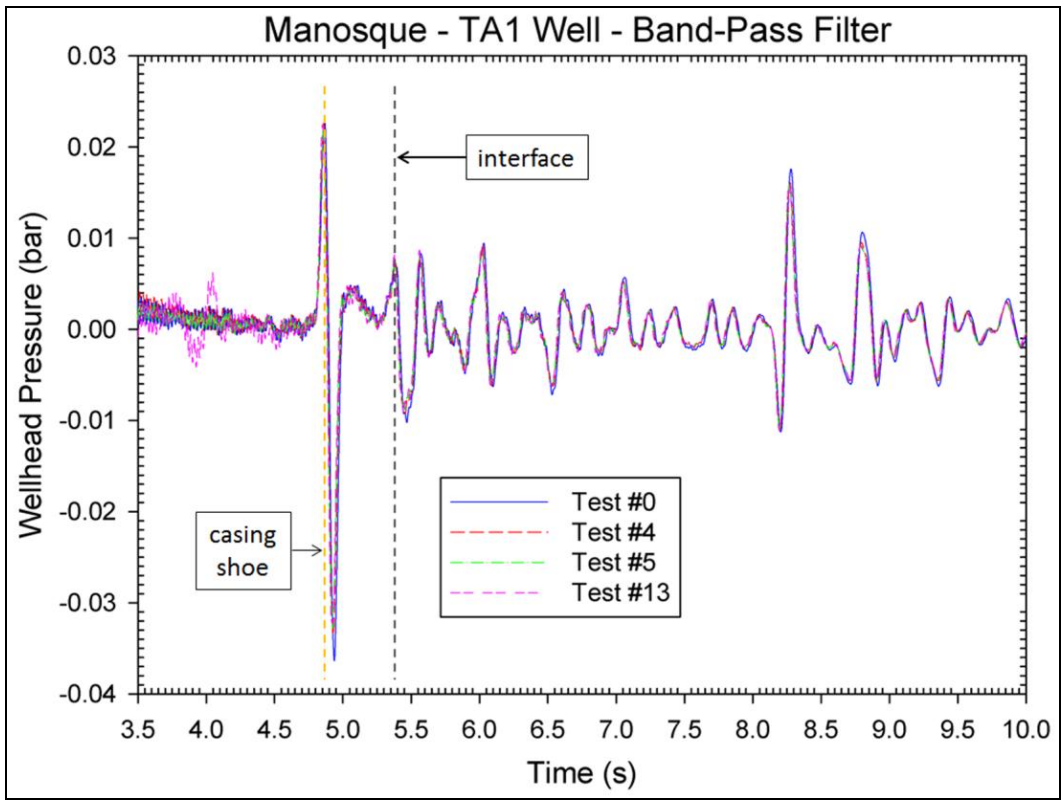

**Figure 14.** Pressure signals filtered from high and low frequencies at Manosque TA1.

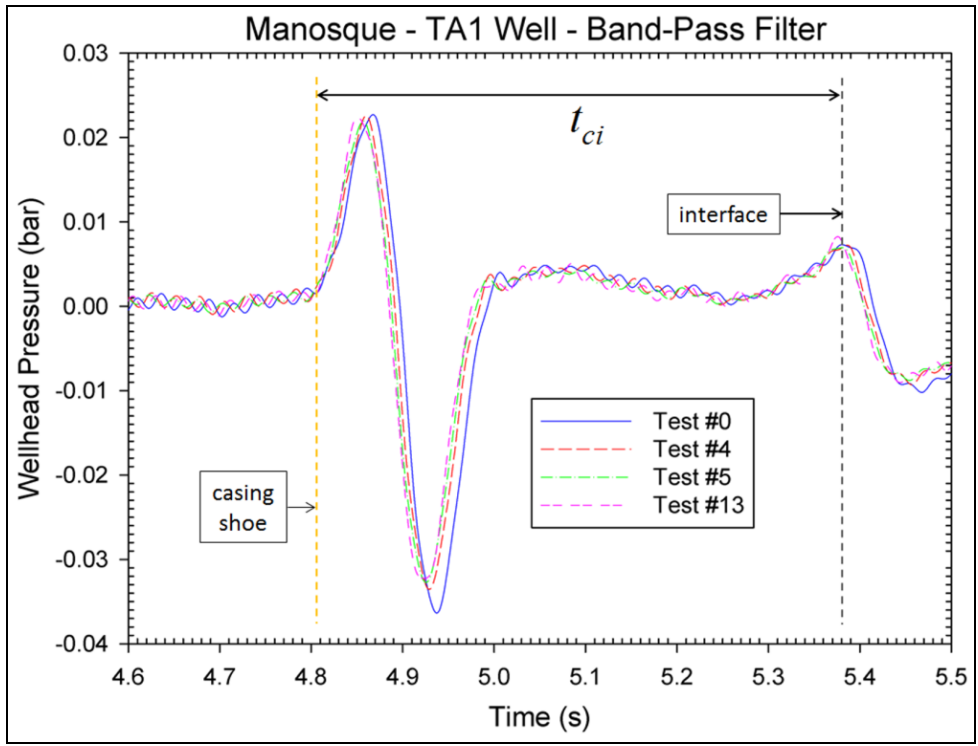

**Figure 15.** Pressure signals filtered from high and low frequencies (close-up) at Manosque TA1.

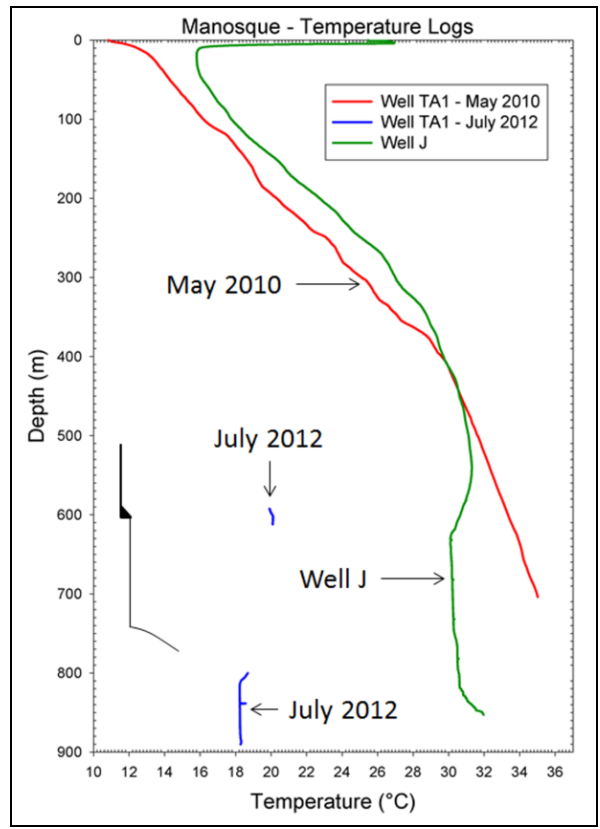

**Figure 16.** Temperature logs in wells TA1 and J.

*4.4 Benchmark — Comparison with downhole logging*

A logging operation was performed by Flodim immediately after the water-hammer tests. Figure 17 shows the CCL Log that was performed first to verify the casing-shoe depth at 603 m. Then, an NPT Log was achieved. The Pulsed Neutron Tool (PNT) records the gamma-Ray induced by reactions between neutrons emitted by the tool and its environment. It is equipped with temperature and Gamma-Ray sensors combined with a DSCL (Dual String Collar Locator) which allows an accurate interface depth determination relative to a cemented casing shoe. The accuracy on the interface depth using NPT is estimated to be in the order of 0.1 m.

The PNT Log performed in TA1 found the nitrogen/brine interface at a depth of 705.1 m (Figure 18) — i.e., very close to the estimation made from the water-hammer analysis (706 m) and inside the uncertainty interval.

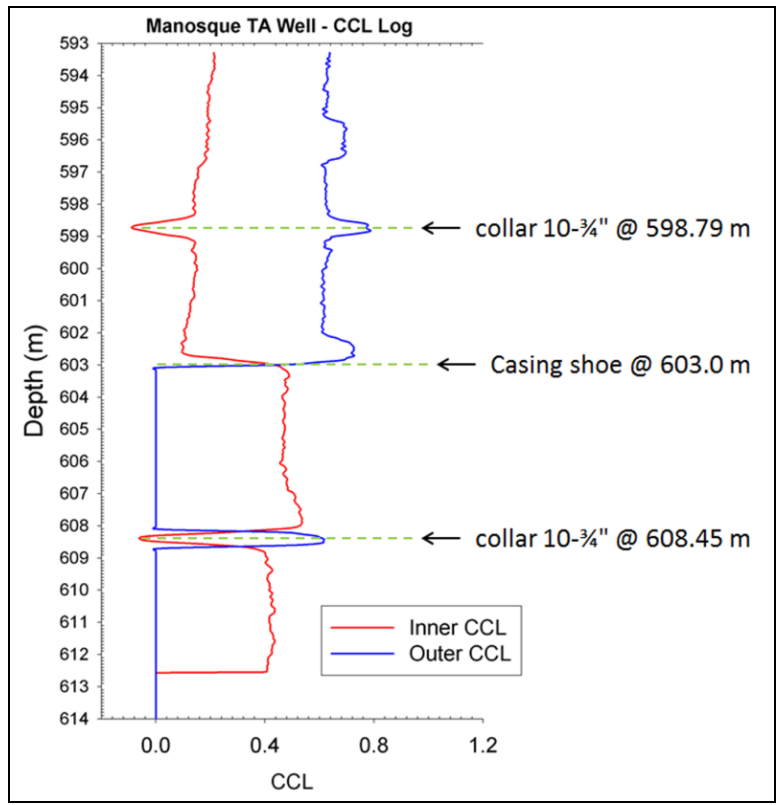

**Figure 17.** CCL Log used to verify the casing-shoe location.

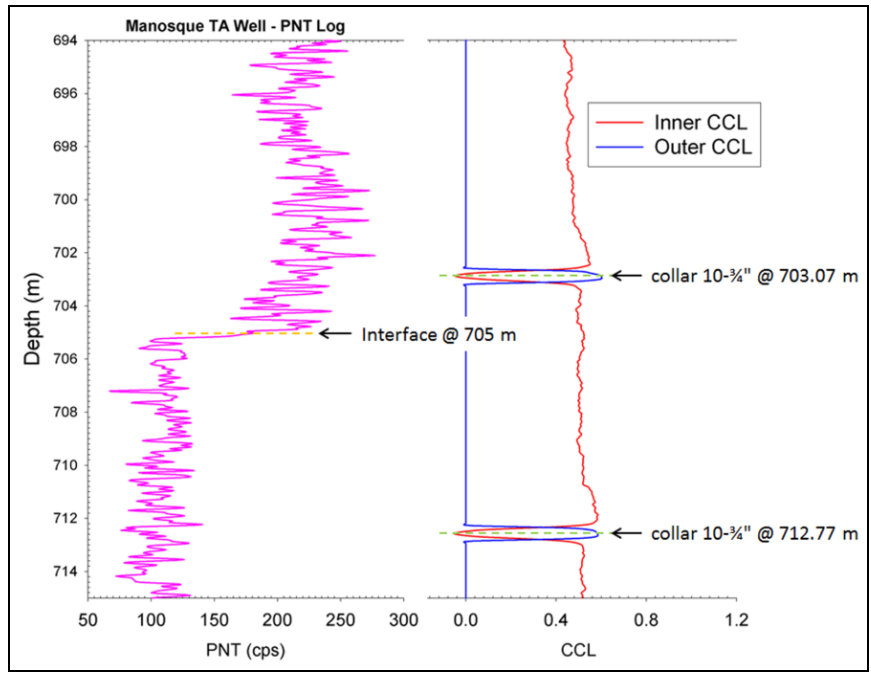

**Figure 18.** PNT Log to verify the interface depth.

### 4.5 Fitting the pressure signal — Chimney profile

A fitting of the wellhead pressure evolution, as recorded during Test #9 (an example of a quick fitting is shown in Figure 19), is possible using the signal from the quartz gauge and the developed software, as was done for the Vauvert PA31 well (see Figure 6). Figure 19 shows only a quick fitting.

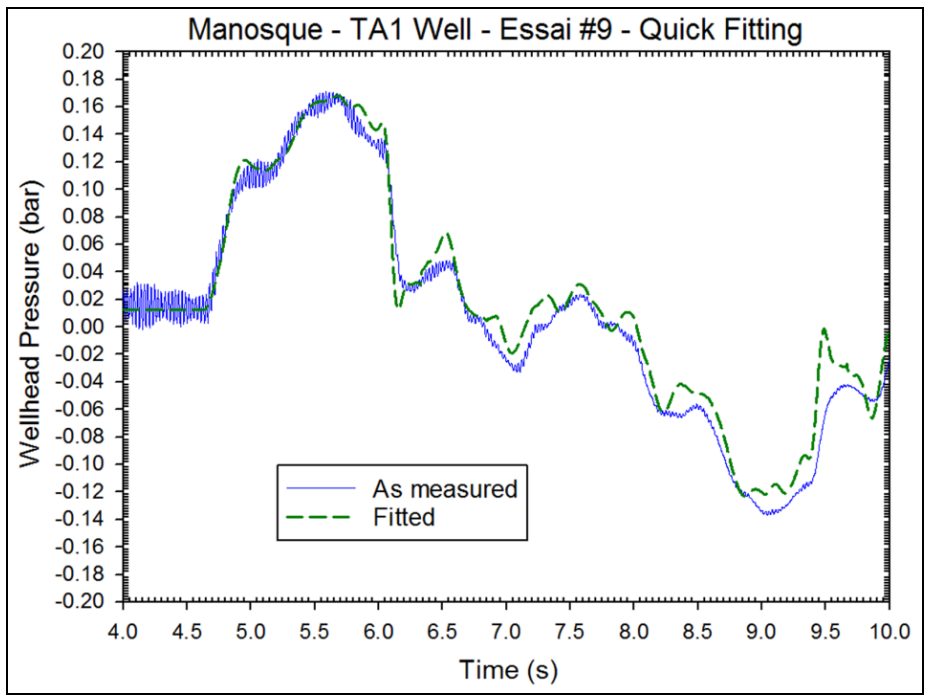

**Figure 19.** Comparison between measured and computed pressure evolutions after fitting at Manosque.

In the case of a gas-filled chimney, this optimization could allow for back-calculation of the vertical profile of the chimney (Figure 20). Further developments are needed to confirm this statement and to compare the back-calculated profile to a direct measurement.

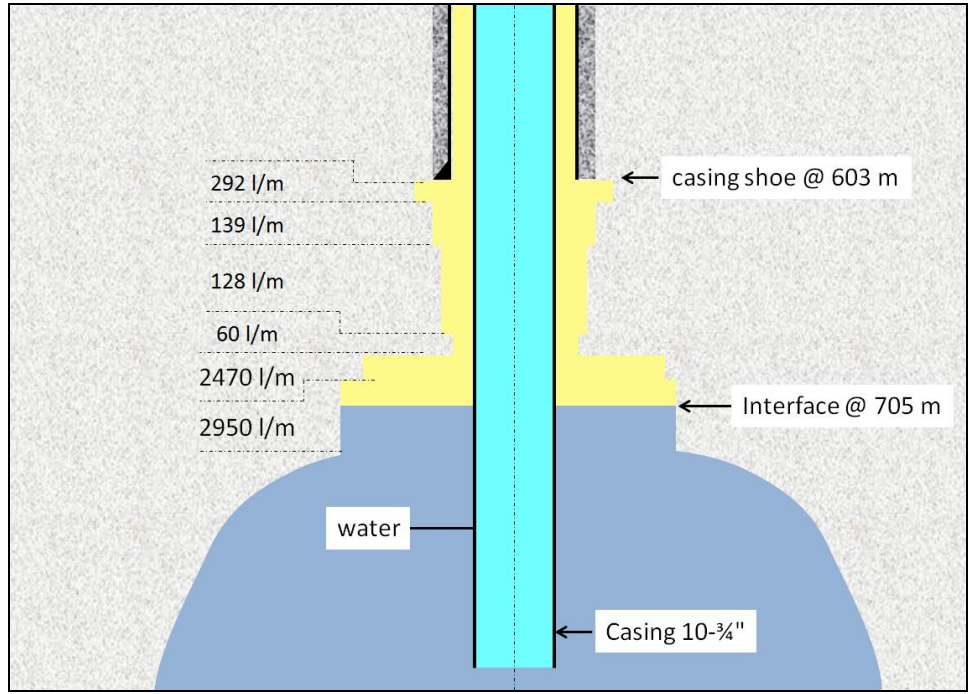

**Figure 20.** Quick back-calculation of the profile of the chimney at Manosque TA1.

## 5. Conclusions

Water hammers commonly are observed in salt cavern wells. They often are considered as a potential hazard to be avoided. Nevertheless, small water hammers provide valuable information about the well. A low-cost and non-intrusive system to record such information has been developed by Brouard Consulting and Ecole Polytechnique. This system includes dedicated software for recording and analyzing water hammers. This device has been installed and successfully tested at the Vauvert brine field to detect any damaged, plugged — or even lost — casing.

The same system has been tested successfully to determine the depth of a gas/brine interface in a chimney during leaching when a gas blanket is used. When compared to other logging tools, the accuracy of this method appears satisfactory. It allows real-time measurement and can be operated remotely. Another interesting feature of this method is that it might allow back-calculation of chimney shape.

**Acknowledgements**

The authors are deeply indebted to Geostock, especially Bruno Colcombet and Xavier Boivinet, and to the Geosel staff at the Manosque facility, whose help in preparing and setting up the system was invaluable.

*Appendix — Helmholtz resonator*

*Theoretical background*

In the presented paper, we analyzed waves generated by a tiny withdrawal of nitrogen at the wellhead that travels through the well in the annular space. An oscillatory phenomenon of another kind intervenes: the interface experiences small movements of long period, and the system behaves as a Helmholtz resonator. The study of these oscillations allows back-calculation of the interface location but, conversely, the oscillations can blur the exact location of the interface, making its measurement by a logging tool more difficult.

Let $h$ be the equilibrium interface depth (Figure 21); $H_{roof} - h$ is the brine-column height in the annular space. As long as the oscillations are relatively rapid, the gas pressurizations/depressurizations can be considered adiabatic — i.e., absolute gas pressure ($P_g$) and gas-column volume, $V_g = H_{cs}S + (h - H_{cs})\Sigma$ are related by the adiabatic relation $P_g V_g^{\gamma} = \text{constant}$, where $\gamma$ is the adiabatic constant. In other words, when small interface displacements are considered,

$$\dot{P}_g = -\left(\gamma \Sigma P_g / V_g\right)\dot{h} \qquad (6)$$

where $S$, $\Sigma$ and $\Sigma'$ are the annular cross-sections above the casing shoe, at interface depth and at cavern roof, respectively.

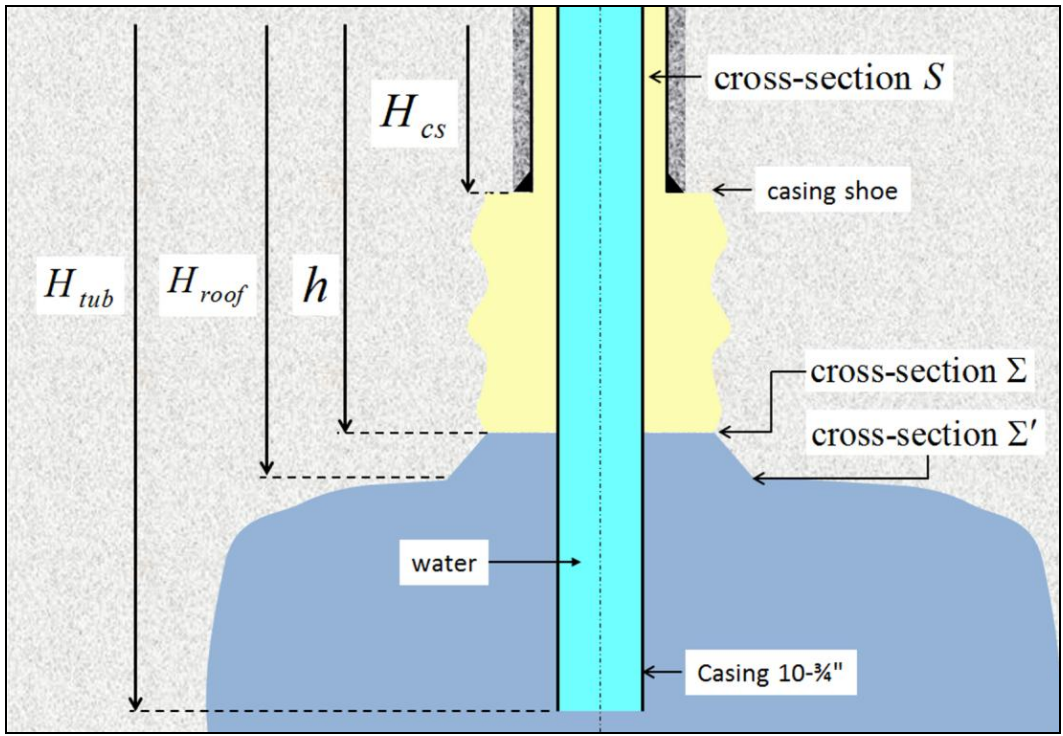

**Figure 21.** Sketch for the Helmholtz resonator calculation.

Cavern pressure changes ($\dot{P}_b$) are related to brine/gas interface changes ($\dot{h}$) and cavern compressibility ($\beta V$):

$$\beta V \times \dot{P}_b = \Sigma \dot{h} \tag{7}$$

The differential equation satisfied by flow rate $Q = \Sigma \dot{h}$ is reached simply by considering that brine in the annular space, with mass $m_b = \rho_b \bar{\Sigma}(H_{roof} - h_o)$ [where $\rho_b$ is brine density, $h_0$ is the initial depth of the interface, $\bar{\Sigma} = (\Sigma + \Sigma' + \sqrt{\Sigma\Sigma'})/3$, and acceleration $\dot{Q}$] is pushed upward by the cavern pressure excess and downward by gas-pressure excess:

$$m_b \dot{Q}/\Sigma = -\left(\frac{\gamma P_g}{V_g} + \frac{\rho_b g}{\Sigma} + \frac{\Sigma'}{\beta V}\right) Q \tag{8}$$

The solution of such a differential equation is a periodic function whose pulsation is

$$\omega^2 = \frac{\gamma \Sigma^2 P_g / V_g + \rho_b g \Sigma + \Sigma \Sigma' / \beta V}{\rho_b \bar{\Sigma}(H_{roof} - h_o)} \tag{9}$$

The corresponding oscillation period is $T_h = 2\pi/\omega$, and the frequency of the Helmholtz resonator is $F_h = \omega/2\pi$.

In the case of a very large roof cross-section compared to the chimney, $\Sigma' \gg \Sigma$, the pulsation can be simplified as follows:

$$\omega^2 = \frac{3\tilde{\Sigma}}{\rho_b (H_{roof} - h_0) \beta V} \quad \text{where} \quad \tilde{\Sigma} = \frac{\Sigma \Sigma'}{\Sigma' + \sqrt{\Sigma\Sigma'}} \tag{10}$$

*Helmholtz — TA1 evidence*

In order to detect this type of oscillation, the pressure gauge was set at the top of the 10-¾" casing. A water hammer then was created at the top of the annular space by withdrawing a tiny amount of nitrogen. Figure 22 shows the brine-pressure evolution recorded 2 minutes after the water-hammer event started. A close-up at the beginning of the water-hammer event shows that generated wave is almost immediately transmitted, at the wellhead, to the inner-string water through the steel tube.

The power spectrum of water pressure oscillations in the 10-¾" casing (Figure 24) clearly shows a quarter-wave (Bérest et al., 1999) with a frequency $F_{qw} = \bar{c}_w / 4 H_{tub}$, where $\bar{c}_w$ is the average celerity of the wave traveling in the water column. As $F_{qw} \approx 0.37$ Hz and $H_{tub} = 890$ m, one obtains $\bar{c}_w \approx 1317$ m/s, which is typical of a sound wave travelling in water-filled tubing. Accurate measurement of this frequency is an easy way to detect any damage or plug in the 10-¾" casing (Brouard et al., 2012).

A smaller frequency, $F_h \approx 0.021$ Hz, corresponding to a period $T_h = 1/F_h = 47.7$ s, also can be observed in the spectrum. To attribute this frequency to the Helmholtz resonator, the horizontal cross-section, $\Sigma'$, at the cavern top must be significantly larger than cross-section $\Sigma$ at interface depth.

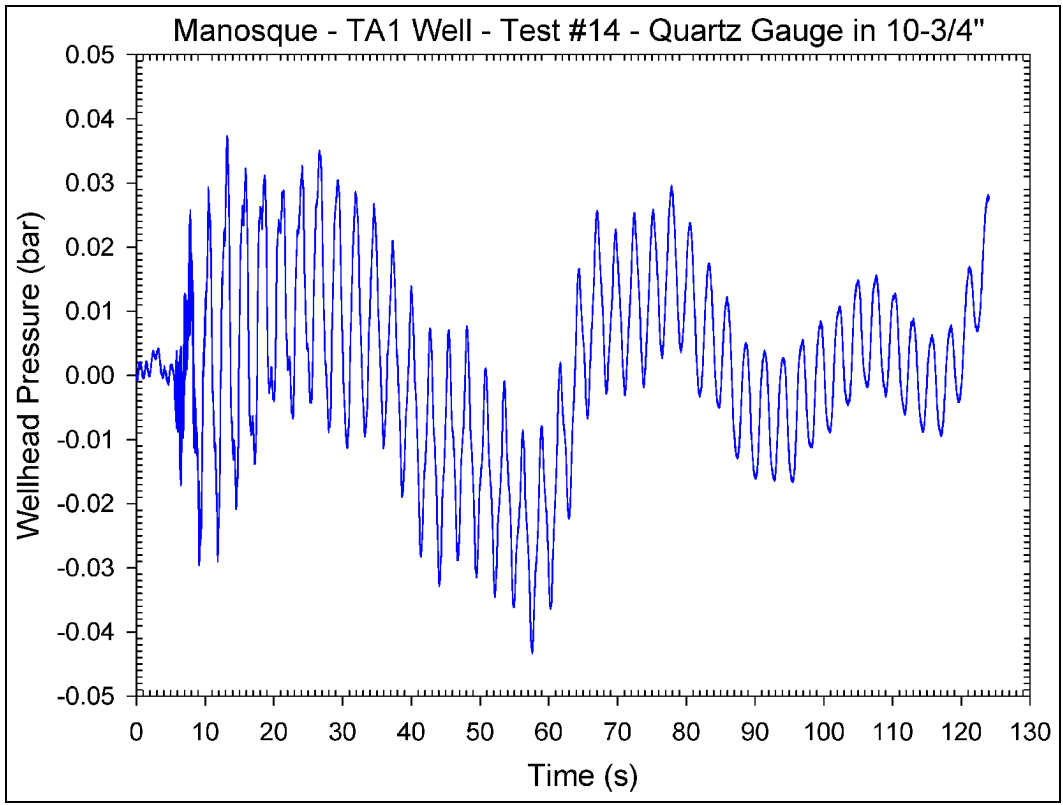

**Figure 22.** Evolution of wellhead brine pressure in the central tubing.

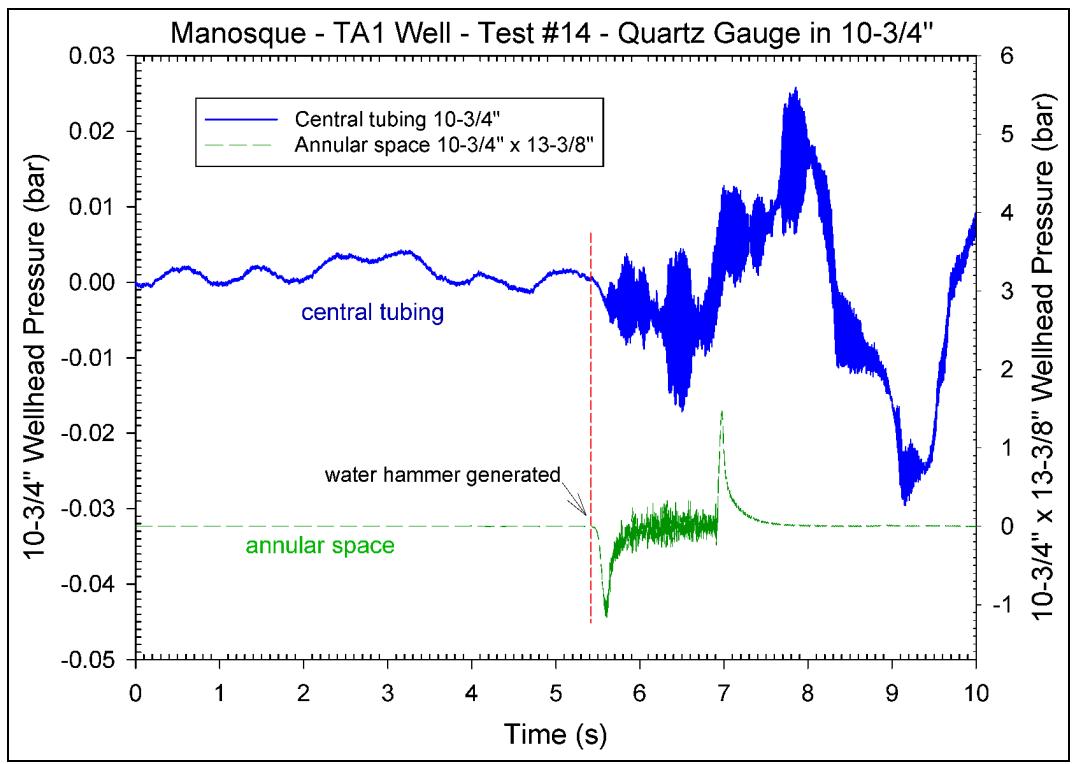

**Figure 23.** Wellhead pressure evolutions at the beginning of the water-hammer event.

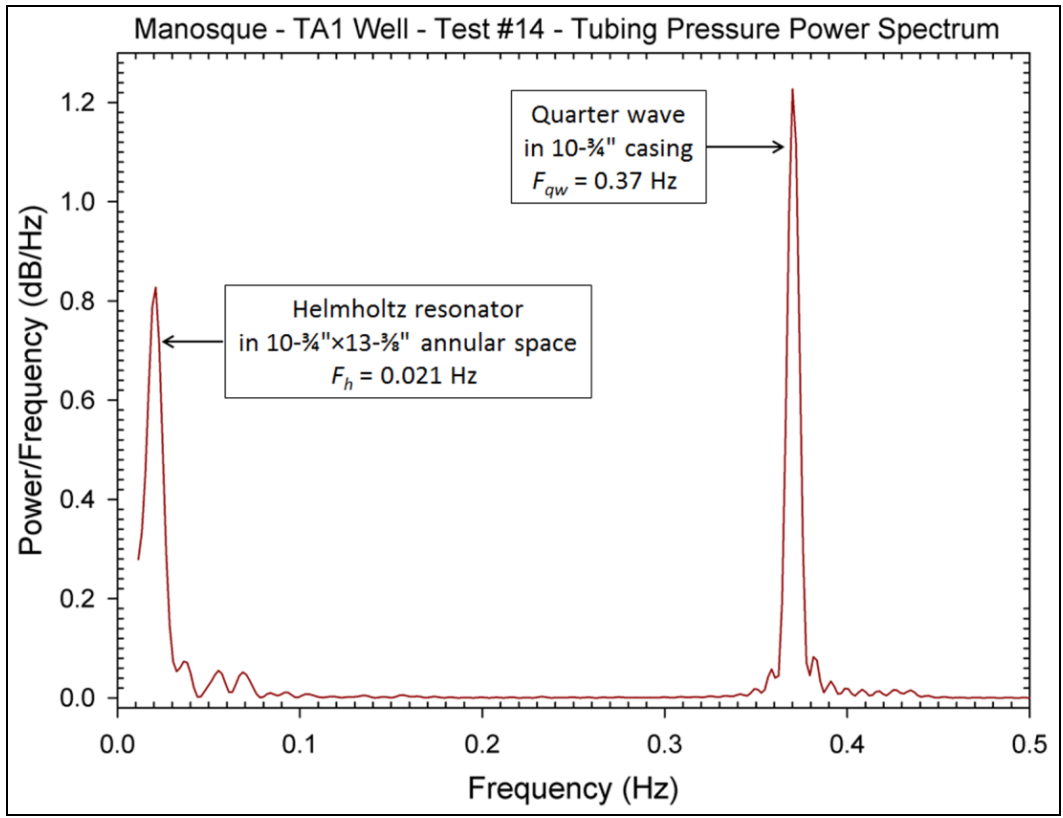

**Figure 24.** Power spectrum of pressure oscillations in the water string.